\documentclass[12pt]{article}

\usepackage{bbold}
\usepackage{amssymb}
\usepackage{graphics}
\usepackage{epsfig}
\usepackage{amsmath,amsfonts}

\def\12{\frac{1}{2}}

\def\Real{\mathbb R}
\def\Zint{\mathbb Z}

\def\be{\begin{equation}}
\def\ee{\end{equation}}
\def\beq{\begin{equation}}
\def\eeq{\end{equation}}
\def\bea{\begin{eqnarray}}
\def\eea{\end{eqnarray}} 
\def\beqa{\begin{equation}\begin{array}{l}}
\def\eeqa{\end{array}\end{equation}}

\def\eqn#1{(\ref{#1})}

\def\eqref#1{eq.~(\ref{eq:#1})}

\def\nn{\nonumber}

\begin{document}
\thispagestyle{empty}

\vspace{.8cm}
\setcounter{footnote}{0}
\begin{center}
{\Large{\bf
Milne and Torus Universes Meet
    }\\[10mm]}

{\sc 
A. Waldron
\\[6mm]}

{\em\small Department of Mathematics, University
of California,\\
Davis, CA 95616, USA\\
{\tt wally@math.ucdavis.edu}}\\[10mm]

{\it Invited contribution to Deserfest: A celebration
of the life and works of Stanley Deser}\\[3mm]
{\it University of Michigan, Ann Arbor, April 2004}\\[10mm]


\bigskip

\bigskip

{\sc Abstract}\\

\end{center}

{\small
\begin{quote}

Three dimensional
quantum gravity with torus universe, $T^2\times\Real$, topology is reformulated
as the motion of a relativistic point particle moving in an  
$Sl(2,\Zint)$ orbifold of flat Minkowski spacetime. The latter is
precisely the three dimensional Milne Universe studied recently by Russo
as a background for Strings. We comment briefly on the dynamics and
quantization of the model.

\end{quote}

\newpage





\begin{quote}

{\it
Stanley has made many fundamental contributions to our understanding
of life in various dimensions. Notable amongst these is dimension three.
Therefore, on the occasion of his seventy third birthday, I present him
a reformulation of three dimensional gravity in terms of a point particle
moving in a flat three dimensional spacetime.
}

\end{quote}

\section{Introduction}

The limited tractability of quantum gravity means that 
minisuperspace reductions
to quantum mechanical degrees of freedom are an important calculational tool.
Moreover in three dimensions, where gravitons carry
no field theoretic degrees of freedom, a minisuperspace reduction might
be expected to faithfully represent the full theory. Indeed, at least
for toroidal spatial topologies\footnote{
Modulo questions of inequivalent quantizations discussed 
by Carlip\cite{Carlip:1995zj}.}, this is the 
case\cite{Martinec:1984fs}:
on any given toroidal spatial slice, a Weyl transformation brings
the 2-metric to a flat one. The space of conformally flat metrics on 
the torus is parameterized by the coset $Sl(2,\Real)/SO(2)$. Furthermore,
a peculiarity of three dimensions is that the space dependence of this
Weyl transformation can be gauged away. Therefore, as observed long ago by Martinec\cite{Martinec:1984fs}, all that remains is the quantum mechanical 
evolution of metric moduli in the space $\Real_+\times Sl(2,\Real)/SO(2)$.
In our previous work\cite{Pioline:2002qz} we made the additional observation
that this system was in fact a relativistic particle with mass
proportional to the cosmological constant. In addition we showed that
this model enjoys a conformal (spectrum generating) symmetry
and falls into a large class of novel conformal quantum mechanical 
models. In this short note we show that for the case of three dimensional
spatially toroidal gravity, the metric moduli space is an $Sl(2,\Zint)$
orbifold of flat three dimensional Minkowski spacetime. 
The derivation of this model is given in Section~\ref{approx}.
Dynamics and solutions are discussed in Section~\ref{dynamics}
while quantization and a discussion of possible physical computations
may be found in Section~\ref{quant}.

\section*{Acknowledgments}

The work presented here is based largely on joint work with Boris Pioline
published in\cite{Pioline:2002qz}. A more detailed examination of quantum 
generalizations is in preparation with Danny Birmingham. 
This work was supported in part by NSF grant PHY01-40365.
It is a pleasure to thank Sergei Gukov for discussions and the Max Planck
Institut f\"ur Mathematik Bonn for hospitality.
Finally, Mike Duff, Jim Liu, Kelly Stelle and Richard Woodard
deserve sustained applause for organizing a terrific celebration
of Stanley's wide ranging scientific achievements which have guided 
not only the direction but style of modern theoretical physics.

\section{Minisuperspace}

\label{approx}

Our model is obtained by rewriting three dimensional cosmological gravity
in the limit where all spatial derivatives are discarded,
so we parameterize the metric as
\be
ds^2=-N(t)^2 dt^2+h_{ij}(t) dx^idx^j\, ,
\ee
where $i,j=1,2$. We study a toroidal topology $T^2\times\Real$ with
$0\leq x^i< 1$. The extrinsic curvature is simply 
\be
K_{ij}=-\frac{1}{2N}\ \dot h_{ij}\, ,
\ee
which implies 
$K\equiv h^{ij}K_{ij}=-\frac{1}{2N}\frac{d}{dt}\log h$ 
with $h\equiv \det h_{ij}$.
The Einstein Hilbert action now reads
\be
S=\int dt\ d^2x\sqrt{h}\ N\ \Big[-K^2+K_{ij}K^{ij}-2\Lambda\Big]\, .
\ee
Extracting the spatial volume 
\be
h_{ij}=h^{1/2}\ \widehat h_{ij}\, ,
\ee
the unit volume metric $\widehat h_{ij}$ is parameterized
by the coset $Sl(2,\Real)/SO(2)$. 
Requiring, in addition, that spatial sections are toroidal, means
that we must identify metrics related by the left action of $Sl(2,\Zint)$. 
The resulting 
$Sl(2,\Zint)\backslash Sl(2,\Real)/SO(2)$ orbifold is just the usual upper 
half plane modded out by modular transformations.
Introducing coset coordinates
$U^M$ ($M=1,2$) and invariant metric
\be
\dot U^M G_{MN}\dot  U^N\equiv -\frac12\ \dot h_{ij}\ \dot h^{ij}
\ee
as well as the field redefinition
\be
\eta\equiv RN\, ,\qquad R\equiv h^{1/2}\, , 
\ee
yields 
\be
S=\frac12\ \int dt\ \Big\{\ 
\frac{1}{\eta}\ \Big[
-\dot R^2+R^2 \dot U^M G_{MN}\dot  U^N
\Big]-4\Lambda \eta
\Big\}\, .
\label{S}
\ee
This is the action of a relativistic particle in three dimensions, 
$(\mbox{mass})^2=4\Lambda$ 
(tachyonic in AdS!) 
with metric
\be
ds_{\mathfrak M}^2=-dR^2+R^2\ dU^M G_{MN} dU^N\, .
\ee
Hence, the time evolution of the metric moduli $U^M$ and $R$
in this minisuperspace truncation is described by the dynamics of
a fictitious particle moving in a three dimensional Lorentzian
spacetime which we will denote~${\mathfrak M}$. 

\section{Geometry of the Metric Moduli Space}

\noindent
To understand the space ${\mathfrak M}$ better, write the
unit volume spatial metric in terms of a zweibein 
\be
\widehat h_{ij}=(e e^t )_{ij}
\ee
where the $Sl(2)$ valued zweibein $e$ has Iwasawa decomposition
\be
e=
\begin{pmatrix}
1&\tau_1&\\&1
\end{pmatrix}
\begin{pmatrix}
\sqrt{\tau_2}&\\&1/\sqrt{\tau_2}
\end{pmatrix}
\ee 
so that 
\be
ds_{\mathfrak M}^2=-dR^2+R^2\ \frac{d\tau_1^2+d\tau_2^2}{\tau_2^2}\, .
\ee
This space is the three dimensional Milne Universe.
Here $-\infty<\tau_1<\infty$, $0<\tau_2$ and $0<R$. Now change coordinates
\be
R=\sqrt{Z^2-X^2-Y^2}\, ,\qquad
\tau_1=\frac{Y}{Z-X}\, ,\qquad
\tau_2=\frac{R}{Z-X}\, .
\ee 
The metric becomes the flat three dimensional Minkowski one
\be
ds^2_{\mathfrak M}=-dZ^2+dX^2+dY^2
\ee
and our fictitious particle action is simply
\be
\label{action}
S=\frac12\ \int dt\ \Big\{\ 
\frac{1}{\eta}\ \Big[
-\dot Z^2+\dot X^2+\dot Y^2\ 
\Big]-4\Lambda \eta
\Big\}\, .
\ee
To complete the discussion, we still need to identify the 
topology of the space $\mathfrak M$. 
Firstly note that the inverse of the above coordinate transformation
is
\be
X = \frac{R}{2}\ \Big(\frac{\tau\overline\tau}{\tau_2}
                    -\frac{1}{\tau_2}\Big)\, ,\quad
Y = R\ \frac{\tau_1}{\tau_2}\, ,\quad
Z = \frac{R}{2}\ \Big(\frac{\tau\overline\tau}{\tau_2}
                    +\frac{1}{\tau_2}\Big)\, .
\ee
(The second solution with an overall $-$ is ruled out
by positivity of $R$ and $\tau_2$ which requires $Z>X$.)
Surfaces of constant $R$ are hyperboloids in the forward 
lightcone (since $Z>X$) isomorphic to the upper half plane~$\mathbb H$.
This is simply Lobacevskii's {\it ur} non-Euclidean geometry depicted
in Figure~\ref{hyperbola}.
\begin{figure}
\begin{center}
\epsfig{file=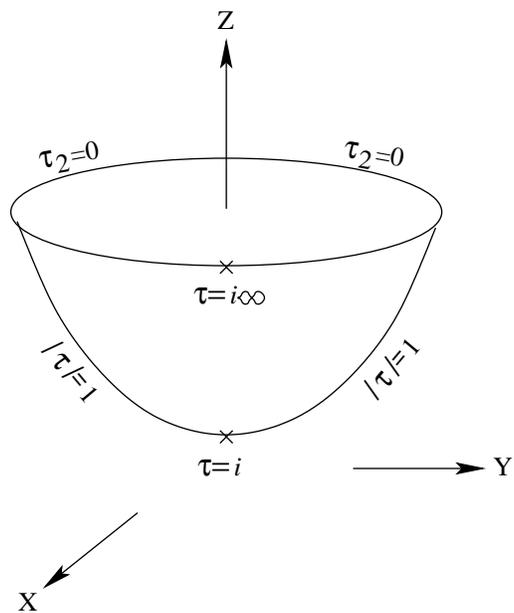,height=8cm}
\end{center}
\caption{Lobacevskii's non-Euclidean plane realized as a
hyperboloid in the forward lightcone.\label{hyperbola}}
\end{figure} 
The boundary $\partial\mathbb H$ at $\tau_2=0$ maps to the boundary of 
each hyperboloid while the cusp at $\tau=i\infty$ maps to $(X,Y,Z)=(\infty,
0,\infty)$. The unit circle $|\tau|=1$ corresponds to the line $\{X=0\}\cap
\{R^2=Z^2-X^2-Y^2\ | \ Z>0\}$. 
Therefore, before modding out by modular transformations, the metric moduli
space is simply the interior of the forward lightcone in three dimensions
with usual flat Minkowski metric.

Now 
we must mod out the upper half plane $PSl(2,\Real)/SO(2)={\mathbb H}$
valued torus moduli $U^M$ by the left action
$PSl(2,\Zint)$ which acts independently on each hyperboloid.
This group is a discrete subgroup of the three dimensional Lorentz group
generated by isometries
$T:\tau\mapsto\tau+1$ and $S:\tau\mapsto-1/\tau$.
In terms of the new variables $(X,Y,Z)$  
these isometries leave $R$ invariant and act as
\bea
T:(X,Y,Z)&\mapsto&\frac12 
\Big(X+2Y+Z , -2[X-Y-Z] ,
            -X+2Y+3Z\Big)\, , \nn\\
S:(X,Y,Z)&\mapsto&(-X,-Y,Z)\, .
\eea
It is easy to verify that an infinitesimal transformation 
$\tau\rightarrow\tau+t$ is generated by
\be
\frac{\partial}{\partial \tau_1}=
Y\partial_X-X\partial_Y+Y\partial_Z+Z\partial_Y
=M_{YX}+M_{YZ}\, ,
\ee
the sum of a rotation in the $(X,Y)$-plane and an $(Y,Z)$-boost
(or better, a lightcone boost). 
The resulting orbifold is precisely the three dimensional
Milne universe studied by Russo\cite{Russo:2003ky} in a String theoretic
context.

Finally, note that the generators of the $Sl(2,\Real)$ isometry subgroup
are
$$
e_+=\partial_{\tau_1}=M_{YZ}+M_{YX}\, ,\quad
h=2\tau_1\partial_{\tau_1}+2\tau_2\partial_{\tau_2}=-2M_{XZ}\, ,
$$
\be
e_-=-\tau_1(\tau_1\partial_{\tau_1}+2\tau_2\partial_{\tau_2})+
\tau_2^2\partial_{\tau_1}=M_{YZ}-M_{YX}\, .
\ee
Hence, we may also view this subgroup as the $SO(2,1)$ Lorentz group
in the natural way.

\section{Dynamics}

\label{dynamics}

The time coordinate $t$ in the relativistic particle model~\eqn{action}
plays no preferred {\it r\^ole}, since the the ``einbein'' $\eta$
ensures reparameterization invariance. Instead classically, we may only predict
trajectories in the space ${\mathfrak M}$. The beauty of this
model is that for a free relativistic particle these are simply straight lines
with slopes subject to Einstein's relativistic dispersion relation.

Explicitly, in the gauge $\eta=1$, straight line geodesics are
\be
Z=Z_0+P_Zt\, ,\quad
X=X_0+P_Xt\, ,\quad
Y=Y_0+P_Yt\, ,
\ee
subject to a mass-shell condition
\be
-P_Z^2+P_X^2+P_Y^2=4\Lambda\, .
\ee
To convert these to metric solutions, note that
the two-metric takes the compact form
\be
\Big(h_{ij}\Big)=
\begin{pmatrix}
Z+X&Y\\Y&Z-X
\end{pmatrix}\, .
\ee
Some fundamental solutions include:

\subsubsection*{Kasner}
Setting $Y(t)=0$ and $\Lambda=0$
the mass shell constraint becomes
\be
(\dot Z+\dot X)(\dot Z-\dot X)=0\, .
\ee
The Kasner solution to this constraint is $Z=t+\frac12$, $X=t-\frac12$
which yields the metric $ds^2=-\frac1{2t}dt^2+2t(dx^1)^2+(dx^2)^2$.
The standard Kasner metric is obtained by changing time coordinates
to ``cosmological time'' $\tau\equiv R=\sqrt{Z^2-X^2}$ so that
\be
ds^2=-d\tau^2+(\tau dx^1)^2+(dx^2)^2\, ,
\ee
which amounts to the gauge $\eta=\tau$ in the relativistic particle model.

\subsubsection*{de Sitter}
Reintroducing the cosmological constant alias the relativistic 
mass $2\sqrt{\Lambda}$ we solve the mass shell constraint via
$Z=2\sqrt{\Lambda}t$, $X=Y=0$ yielding metric
$ds^2=-\frac{1}{4\Lambda t^2}dt^2+2\sqrt{\Lambda}t\ [(dx^1)^2+(dx^2)^2]$. 
Changing time coordinates $\tau=\tau(t)$ to the gauge 
$\eta=\exp\Big(2\sqrt{\Lambda}\tau\Big)$ 
yields the steady state de Sitter metric
\be
ds^2=-d\tau^2+e^{2\sqrt{\Lambda}\tau}[(dx^1)^2+(dx^2)^2]\, .
\ee

\subsubsection*{Anti de Sitter}

We can also consider tachyonic trajectories corresponding
to negative cosmological constant,
$X=2\sqrt{|\Lambda|}t$, $Y=0$ and $Z=Z_0$ (say).
This yields a novel Anti de Sitter metric
\be
ds^2=-\frac{dt^2}{Z_0^2+4\Lambda t^2}+
(Z_0+2\sqrt{|\Lambda|}t)(dx^1)^2 + (Z_0-2\sqrt{|\Lambda|}t)(dx^2)^2\, .
\ee
This metric becomes singular at $t=\pm Z_0/(2\sqrt{|\Lambda|})$
at which points the volume of spatial slices $R=0$. Therefore, it 
represents only a coordinate patch of Anti de Sitter space. It would be
interesting to study the compatibility of the $\Zint^2$ torus orbifold
on the spatial coordinates $(x^1,x^2)$ with the geodesic completion of
the above metric.

\section{Quantization and Conclusions}

\label{quant}

We have presented a simple three dimensional
relativistic particle model that describes toroidal
gravity in three dimensions. The model would be trivially
solvable if it were not for the $Sl(2,\Zint)$ orbifold of
the flat particle background necessary to identify equivalent
gravity metrics. The Hilbert space of physical states in this
model is dictated by the Hamiltonian constraint
\be
\Big\{-\partial_Z^2+\partial_X^2+\partial_Y^2+4\Lambda\Big\}\, \Psi=0\, .
\ee
Here we have a chosen a particular quantization corresponding
to the natural operator ordering stated. An initial investigation of
this Klein-Gordon equation has been conducted 
in\cite{Russo:2003ky}, the key difficulty
being to automorphize with respect to $Sl(2,\Zint)$. 
A more detailed study will appear\cite{Birmingham}.

Finally, an old observation\cite{Rubakov:1988jf} is that Hamiltonian 
constraints taking this
Klein-Gordon form naturally imply a second quantization of the theory,
jocularly dubbed ``third quantization''. A natural candidate for interactions
would be 
$\phi^3$ theory in three dimensions. Automorphic particle scattering amplitudes
would then amount to probabilities of cobordisms with torus boundary
in three dimensional quantum gravity.

%
%
%
%


\begin{thebibliography}{0}

\bibitem{Martinec:1984fs}
E.~J.~Martinec,
{\it Phys.\ Rev.\ D} {\bf 30}, 1198 (1984).

\bibitem{Carlip:1995zj}
S.~Carlip,
{\it J.\ Korean Phys.\ Soc.}\  {\bf 28}, S447 (1995)
[arXiv:gr-qc/9503024].

\bibitem{Pioline:2002qz}
B.~Pioline and A.~Waldron,
{\it Phys.\ Rev.\ Lett.}\  {\bf 90}, 031302 (2003)
[arXiv:hep-th/0209044].

\bibitem{Russo:2003ky}
J.~G.~Russo,
{\it Mod.\ Phys.\ Lett.\ A} {\bf 19}, 421 (2004)
[arXiv:hep-th/0305032].

\bibitem{Birmingham}
D.~Birmingham and A.~Waldron, in preparation.

\bibitem{Rubakov:1988jf}
V.~A.~Rubakov,
{\it Phys.\ Lett.\ B} {\bf 214}, 503 (1988).
S.~B.~Giddings and A.~Strominger,
{\it Nucl.\ Phys.\ B} {\bf 321}, 481 (1989).




\end{thebibliography}
\end{document}